\documentclass[12pt,preprint]{aastex}

\shorttitle{SA in Radio Galaxies} \shortauthors{Fan et al.}

\begin{document}

\title{Stochastic Acceleration in the Western Hotspot of Pictor A}

\author{Zhong-Hui Fan\altaffilmark{1}, Siming Liu\altaffilmark{2},
Jian-Min Wang\altaffilmark{1}, Christopher L. Fryer,\altaffilmark{2,
3} and Hui Li\altaffilmark{2} }

\altaffiltext{1}{Key Laboratory for Particle Astrophysics, Institute
of High Energy Physics, Chinese Academy of Sciences, 19B Yuquan
Road, 100049 Beijing, China; zhfan@ihep.ac.cn, wangjm@ihep.ac.cn}
\altaffiltext{2}{Los Alamos National Laboratory, Los Alamos, NM
87545; liusm@lanl.gov, hli@lanl.gov} \altaffiltext{3}{Physics
Department, The University of Arizona, Tucson, AZ 85721;
clfreyer@lanl.gov}

\begin{abstract}
{\it Chandra}'s high resolution observations of radio galaxies
require a revisit of the relevant electron acceleration processes.
Although the diffusive shock particle acceleration model may explain
spectra of spatially unresolved sources, it encounters difficulties
in explaining the structure and spectral properties of recently
discovered {\it Chandra} X-ray features in several low-power radio
sources. We argue that these observations strongly suggest
stochastic electron acceleration by magnetized turbulence, and show
that the simplest stochastic particle acceleration model with energy
independent acceleration and escape timescales can overcome most of
these difficulties. We use the bright core of the western hotspot of
Pictor A as an example to demonstrate the model characteristics,
which may be tested with high energy observations.

\end{abstract}


\section{Introduction}
In the classical diffusive shock particle acceleration model the
observed relativistic electrons in radio galaxies are accelerated at
the edges and knots of jets. For relatively more luminous FR II
radio galaxies, electrons may also be energized in radio hotspots,
where jets terminate. It is usually assumed that the jet power is
initially stored in the bulk motion of outflows and is converted
into the internal energy of the jet plasma by shocks at the observed
radio bright features \citep{br74}.
The model requires that particles repeatedly pass the shock front,
characterized by a jump in the density and velocity field, to reach
very high energies. Although the physical processes responsible for
the scattering and/or diffusion of particles through the shock front
in most collisionless astrophysical plasmas are not well
established, it has been widely cited due to its simplicity and its
achievement in explaining the spectra of spatial unresolved sources
\citep[e.g.][]{hbw01}.

It however is challenged by recent high spatial and
spectral resolution observations with {\it Chandra}.
Although nonthermal X-rays from radio galaxies can be produced by
relativistic electrons through synchrotron, synchrotron
self-Comptonization (SSC), and inverse Compton scattering of the
cosmic microwave background radiation (CMBR) or other background
photons, X-rays from jets of FR I galaxies and some low-power
hotspots of FR II galaxies are generally attributed to synchrotron
process \citep{wbh01, h04}. If the magnetic field is nearly in
energy equipartition with the relativistic electrons, the
synchrotron cooling timescale of electrons producing the observed
X-rays is short, and the electrons need to be accelerated very close
to the observed compact X-ray features, which usually are displaced
from peaks of the corresponding radio sources and occasionally do
not have optical and/or radio counterparts \citep{hbw01, hck07, m02,
wy02}. The diffuse X-ray emission from jets of FR I galaxies, on the
other hand, indicates a distributed particle acceleration process.
The X-ray spectral hardening and relatively high
X-ray flux as compared to the extrapolation from optical fluxes of
some compact sources are also puzzling.

The bright core of the western hotspot of Pictor A (WHPA), a famous
nearby\footnote{At a redshift of $z = 0.035$, the luminosity
distance to Pictor A is $d_{\rm L} = 151.9$\,Mpc, and the conversion
scale is $1\arcsec = 0.687$\,kpc for the modern cosmology with
$\Omega_{\rm M} = 0.27$, $\Omega_{\Lambda} = 0.73$, and $H_0 =
71$\,km\,s$^{-1}$\,Mpc$^{-1}$ \citep{s03}.}
FR II galaxy, is an excellent candidate to demonstrate
difficulties of the shock model.
\citet{myr97} discussed its radio to optical spectrum, and
\citet{w01} reported X-ray observations with {\it Chandra}. The
latter studied in detail several possible X-ray emission processes
and found that none of them can explain the broadband spectrum (BBS)
naturally. They concluded that the X-rays are most likely
synchrotron emission by an electron population distinct from these
responsible for the radio to optical spectrum.
In this picture the fact that the radio, optical, and X-ray
emissions come from the same spatial region would have to be a
coincidence.

In this Letter, we show how the simplest stochastic particle
acceleration (SA) model 
explains the BBS of WHPA. The model also predicts a spectral cutoff
near the {\it Chandra} band and an intrinsic volume of $0\farcs03^3$
of the particle acceleration region (AR), which may be tested with
future observations. In \S\ 2, we discuss the major difficulties of
the shock model in explaining {\it Chandra} observations and why the
SA by turbulence is favorable. A simple SA model is built and
applied to WHPA in \S\ 3. In \S\ 4 we draw conclusions and
discuss how the model may also explain other {\it Chandra} X-ray
features.

\section{Shock VS Stochastic  Acceleration}
The synchrotron cooling time of an isotropic electron population is
given by $\tau_{\rm syn} = 9 m_e^3 c^5/4 e^4 B^2\gamma = 24.6
\gamma^{-1}B^{-2}$ yr, where $m_e,\ e,\ \gamma,\ c,$ and $B$ are the
electron mass, charge, Lorentz factor, the speed of light, and
magnetic field, respectively. Here and in what follows, c.g.s. units
are adopted. The synchrotron power spectrum peaks near
$\sim1.7\times 10^{-11} B \gamma^2$ keV. For a $0.1$ mG magnetic
field, typical for jets and hotspots of radio galaxies, the X-ray
emitting electrons have $\gamma>3\times 10^7$ and $\tau_{\rm
syn}<82$ yr. The corresponding light travel distance is $<25$ pc,
which is comparable to or smaller than the {\it Chandra} resolution
for most radio galaxies. The X-ray emitting electrons therefore need
to be accelerated near the observed X-ray sources. In the standard
diffusive shock model, particles are accelerated at a thin shock
front and are carried away by the shocked flow into the downstream.
The simplest 1-D model predicts a power-law synchrotron spectrum
that cuts off at the frequency where the synchrotron cooling time is
equal to the flow travel time from the shock front \citep{hm87}.
Then the spatially integrated spectrum from any segment of the
downstream region can be fitted with a broken power-law, which cuts
off at the cutoff frequency of the boundary closer to the shock
front. The break frequency corresponds to the cutoff frequency of
the farther boundary. The model has difficulties in accounting for
the following {\it Chandra} observations:

{\bf (i)} 
If the diffuse X-ray emission from jets is to be associated with
spatially unresolved shocks, the jet dynamics must be sophisticated
enough to induce many shocks. {\bf (ii)} The X-ray spectral
hardening is observed in a few jet knots of FR I galaxies and
hotspots of FR II sources, e.g., knot WK7.8 of PKS 0637-752
\citep{s00}, some knots in the jets of M87, 3C 120 and NGC 6251
\citep{wy02, hmw04, s04}, WHPA \citep{w01}, and hotspot P1/2 of
3C227 \citep{hck07}.
\citet{da02} suggested that it may be due to the Klein-Nishina
effects of inverse Compton scattering of CMBR by relativistic
electrons that produce the observed X-ray through synchrotron
process. The model requires an extremely low magnetic field with an
energy density comparable to that of the local CMBR or very strong
Doppler effects and may not be applicable to most of the observed
X-ray sources in radio galaxies, which are still quite powerful and
do not show strong Doppler effects. The extreme hardening of hotspot
P1/2 of 3C227 and inner ($<50\farcs$) jet of Centaurus A is even
more challenging. The photon spectral indexes of these sources can
be as small as 1.6 corresponding to an electron spectral index of
2.2, which can not be a cooling spectrum in the shock model. The
lifetime of these sources
therefore must be shorter than the synchrotron cooling time of X-ray
emitting electrons. Given the observed source extension, a thin
shock front may not be able to accelerate enough electrons to
produce the observed X-ray flux over such a short time scale ($<100$
yr). {\bf (iii)} The displacement of X-ray peaks with respect to the
corresponding optical and/or radio peaks \citep{wy02, hbw01}. The
shock model predicts that higher energy emission is produced closer
to the shock front, and the distance of synchrotron emission peak
from the shock front is inversely proportional to the square root of
the emission frequency for a constant downstream velocity and
magnetic field \citep{m89}. There is no observational evidence for
such a source structure. On the contrary, much more complicated
structures are seen in a few low-power hotspots of FR II galaxies
\citep{hck07}.

While radio observations are generally consistent with shock models
for hotspots \citep{m89, car91}, there are indications of electrons
being accelerated directly by magnetic fields in giant radio
galaxies \citep{k04}. Optical detections of some of these radio
sources show that electrons are accelerated in extended regions that
may not be associated with shocks \citep{myr97}. Little is known
about this distributed acceleration process driven by free energy
dissipations. The high Reynolds numbers of the observed radio
sources suggest that turbulence may mediate these dissipations and
accelerate some particles, the so-called SA, a second order Fermi
acceleration that has been ignored in most astrophysical situations
due to its relatively lower efficiency. The observed cutoffs of
synchrotron spectra suggest that the acceleration is less efficient
than that given by the diffusive shock with Bohm diffusion
\citep{bm03, h04}. Moreover, for strongly magnetized plasmas, the
Alfv\'{e}n velocity can be comparable to $c$ and SA can be effective
\citep{pl04}. These combined with the {\it Chandra} observations
strongly suggest that SA may be the dominant acceleration mechanism
in radio galaxies.

SA occurs wherever there are turbulent magnetic fields and can
operate over an extended region with its size dictated by the
turbulence generation and decay rates. Emission from the AR
therefore can be significant. We consider the simplest SA model with
energy independent acceleration ($\tau_{\rm ac}$) and escape
($\tau_{\rm esc}$) times. Then the steady-state particle
distribution in the AR is given by \citep{s84, pp95, lmp06}:
$$
N_{\rm ac}(\gamma)\propto
\gamma^{\delta+2}\exp{(-\gamma/\gamma_{c})}
$$
\begin{displaymath}
\times\left\{ \begin{array}{ll}
\int_0^\infty x^{\delta-1}(1+x)^{3+\delta} e^{-\gamma x/\gamma_{c}}
{\rm d}x \,\, &  {\rm for}\,\, \gamma > \gamma_{\rm inj}  \cr
\int_0^1 x^{\delta-1}(1-x)^{3+\delta} e^{\gamma x/\gamma_{c}} {\rm
d}x \,\, & {\rm for}\,\, \gamma < \gamma_{\rm inj}
\end{array} \right.
\end{displaymath}
where $\delta=({9/4}+{\tau_{\rm ac}/\tau_{\rm esc}})^{1/2}-1.5$,
$\gamma_c =9 m_e^3 c^5/4 e^4 B^2 \tau_{\rm ac}$, and we have assumed
that electrons are injected into the AR at $\gamma_{\rm inj}$ and
the synchrotron cooling dominates.

As in the shock model, electrons escaping the AR may produce
most of the observed emission in a region dominated by cooling
processes. The corresponding particle distribution at time $t$ since
the acceleration starts is given by \citep{bg70}
\begin{eqnarray}
N_{\rm cool}(\gamma, t) = \int_{0}^t {\dot{\gamma}(\gamma_0)\over
\dot{\gamma}(\gamma)}{N_{\rm ac}(\gamma_0) \over \tau_{\rm esc}}{\rm
d} t_0
\nonumber \,,
\end{eqnarray} where  $\dot{\gamma}(\gamma)<0$ is
the energy loss rate and $t_0-t=\int_{\gamma}^{\gamma_0} {\rm d}
\gamma^\prime/\dot{\gamma}(\gamma^\prime)$. Note that if
$t_0-t<\int_{\gamma}^{\infty} {\rm d}
\gamma^\prime/\dot{\gamma}(\gamma^\prime)$, the integrant should be
set to zero. The thin dashed and dotted-dashed lines in Figure
\ref{ne} give $N_{\rm ac}$ and $N_{\rm cool}$, respectively, for
$\gamma_{\rm inj} = 1800$, $\delta = 1.48$, $\gamma_c = 1.04\times
10^7$, $t = 17.9 \tau_{\rm ac}$ (see \S\ \ref{App} on values of
these parameters) and $\dot{\gamma}\propto -\gamma^2$. We see that
the spatially integrated particle distribution as indicated by the
thin solid line is very similar to the result of the shock model,
although the AR can be much more extended in the SA model.

\section{Application to WHPA}
\label{App}

\begin{figure}
\plotone{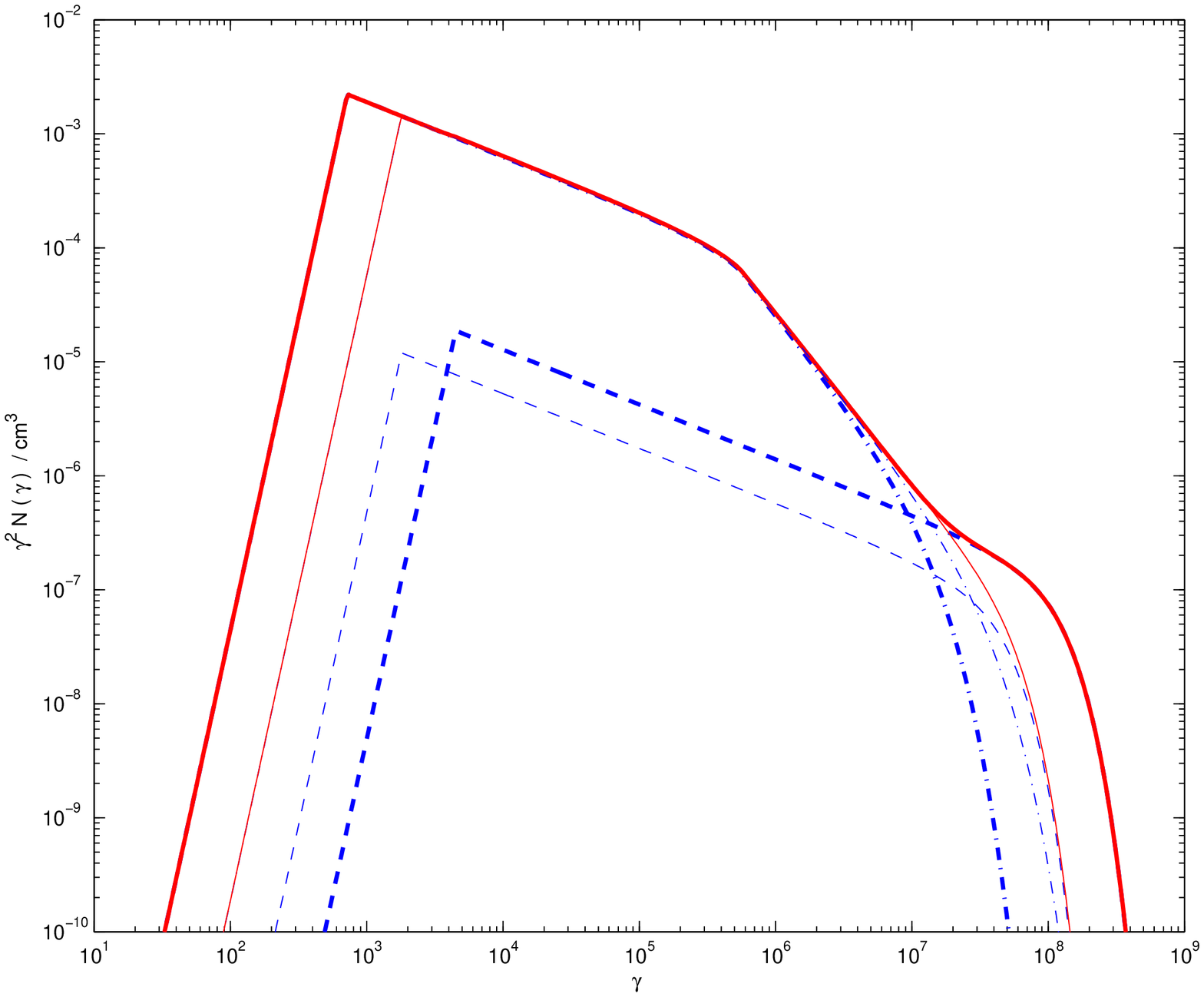} \caption{Effective electron distributions for
$\Delta r=2.5$ (thick) and $1.0$ (thin). The dashed and
dotted-dashed lines are for the acceleration and CR, respectively.
The solid lines are the sums of the two. The normalization is for
the fiducial model. See text for the rest of the model parameters.
\label{ne}}
\end{figure}

To apply the model to hotspots of radio galaxies, one needs to take
into account the effects of adiabatic expansion \citep{car91, bm03}.
Such an expansion is expected since the pressures of hotspot plasmas
are usually much higher than their surroundings. To simplify the
model, we assume that this expansion is decoupled from the
acceleration and cooling processes, i.e. the AR experiences a quick
adiabatic expansion before merging into the cooling region (CR). For
a spherical AR with its radius increased by a factor of $\Delta r$
during the expansion, the magnetic field energy density decreases by
a factor of $\Delta r^4$ and the energy of relativistic electrons
decreases by a factor of $\Delta r$. After the expansion, the plasma
enters a relatively uniform cooling region. We are mostly interested
in the synchrotron emission. To reproduce the optically thin
synchrotron spectrum of the AR with the magnetic field of the CR,
one needs to increase  the electron energy and total number by a
factor of $\Delta r$ and $\Delta r^2$, respectively, since the
synchrotron frequency and luminosity are proportional to $B\gamma^2$
and $B^2\gamma^2$, respectively. The thick lines in Figure \ref{ne}
show these equivalent electron distributions for $\Delta r = 2.5$.
Emission from the adiabatic expansion phase can be ignored as far as
our assumption of quick expansion withheld. We see that the overall
spectrum becomes harder right below the cutoff due to dominance of
the AR there.

\begin{figure}
\plotone{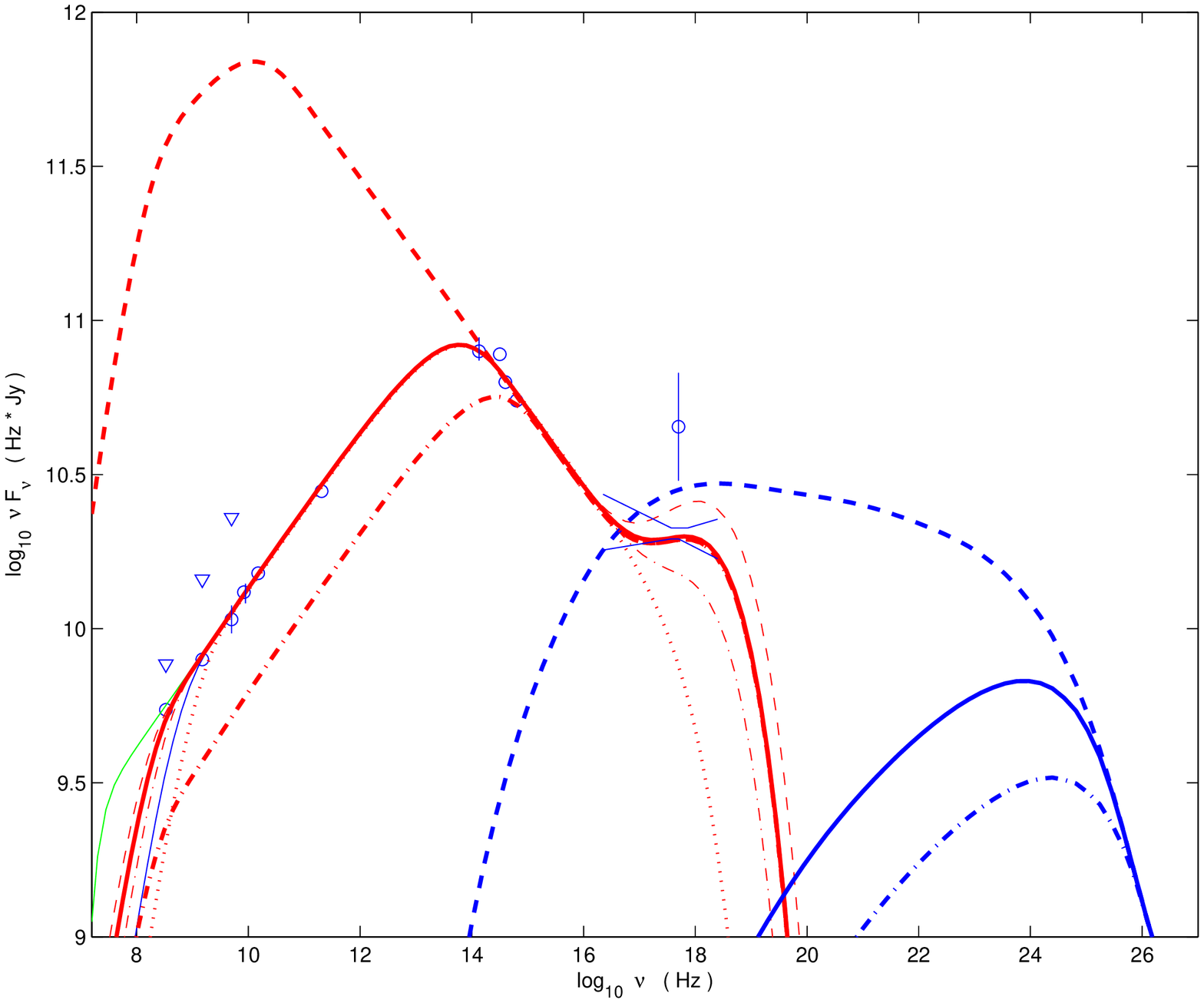} \caption{Best fit to the spectrum of WHPA and
effects of $\Delta r$, $\gamma_{\rm inj}$, and $t$.  The data are
from \citet{w01}. The thick solid line is the best fit with $\Delta
r=2.5$, $B = 85\ \mu$G, $n_e = 2.8\times 10^{-6}$cm$^{-3}$, and
$t$=5.2 kyr. The thin dashed, dotted-dashed, dotted lines are for
$\Delta r=3.0, 2.0, 1.0$, respectively. The two thin solid lines are
for $\gamma_{\rm inj}$=3000, and 10 with the former having a lower
radio flux. The downward triangle radio points are for spatially
unresolved fluxes from \citet{prm97}. Their relatively harder
spectrum suggests a spectral break near $\gamma \sim 1000$. The
higher energy components are produced through SSC. The thick
dotted-dashed and dashed lines are for $t$=2.3 and 400 kyr,
respectively. The SSC component dominates the X-ray flux for the
latter. \label{model}}
\end{figure}

Figure \ref{model} shows the BBS of WHPA, where all data are taken
from \citet{w01}. To fit it with the synchrotron spectrum of the
electron distribution in Figure \ref{ne}, one can have a $\Delta
r>1$ and use the hardening of the electron distribution below the
cutoff to produce the observed X-ray spectral hardening.  For a
given $B$, one can constrain $\gamma_c$ and $\Delta r$ by fitting
the X-ray spectrum.  The radio spectral index $\alpha=0.74$ implies
$\delta=1.48$. Then the frequency of the synchrotron spectral break
from the radio to X-ray band determines $t$. The intrinsic dimension
of the bright core of WHPA is 2$\farcs$ $\times$ 1$\farcs$ $\times$
1$\farcs$ \citep{prm97}, the equivalent radius
$R=0\farcs78=1.65\times 10^{21}$ cm. The normalization of the
electron distribution is then determined by the flux density
(Doppler effects are ignored here). \citet{mig07} find the electron
to magnetic field energy density ratio $U_e/U_B\sim 25(\gamma_{\rm
inj}/50)^{-0.6}$ in the radio lobes of Pictor A. Since relativistic
electrons and magnetic field experience the same adiabatic expansion
loss from hotspots to radio lobes and the former is also subject to
radiative losses, this ratio should be higher in the hotspot. We
assume $U_e/U_B=15(\gamma_{\rm inj}/1800)^{1-\delta}$ and adjust $B$
to fit the BBS. The thick solid line in Figure \ref{model} shows the
corresponding best fit,  where, besides the parameters discussed
above, $B =$85 $\mu$G and the electron density $n_e = 2.8 \times
10^{-6}$ cm$^{-3}$ in the CR. The X-ray spectral hardening is
reproduced with a $\Delta r = 2.5$. The lifetime of the CR $t =5.2$
kyr.  The energy flux carried away by electrons and magnetic field
from the AR is $1.5\times 10^{45}$ erg s$^{-1}$, which is a
reasonable low limit of the jet power. For an electron-proton plasma
without a low energy component, the corresponding Alfv\'{e}n
velocity is $0.26 c$ implying a highly magnetized plasma.

The equivalent radius of the AR $R_a=4.0\times 10^{19}$ cm.
The corresponding light crossing time of 42 yrs is much longer than
the cooling time of 7.4 yr at $\gamma_c$ (i.e. the acceleration time
$\tau_{\rm ac}$) in the AR. Therefore the shock model has
difficulties in reproducing the X-ray flux since the lifetime of the
X-ray emitting electrons (with $\gamma\sim \gamma_c$) is too short
for a thin shock front to generate enough such electrons to
reproduce the observed flux. One can remedy the shock model by
decreasing the magnetic field in the downstream region or increasing
the shock front cross section.
The former requires a magnetic field energy density far below the
equipartitional value with the electrons, which seems unlikely given
that a strong magnetic field is required to convert the jet energy
into the energy of particles.
For a downstream flow velocity $v_d$, the latter requires a cross
section greater than $4\pi R_a^3/3v_d \tau_{\rm ac} \simeq
(0.1c/v_d)(0\farcs3)^2$. From $\delta= 1.48$, we have $\tau_{\rm
esc} = \tau_{\rm ac}/6.6$, the thickness of the AR needs to be on
the order of 1.1 lyr $\sim 0.5$ marcsecond. The corresponding cross
section is $2.5\times 10^{41}$ cm$^2$ $\sim 0\farcs24^2$, which is
smaller than that of the shock model.

The adiabatic expansion phase plays a crucial role in the X-ray
spectral hardening. 
The synchrotron spectra for $\Delta r$=1.0, 2.0, and 3.0, are shown by the
thin dotted, dotted-dashed, and dashed lines in Figure \ref{model},
respectively, 
where $n_e$ is proportional to $\Delta r^{\delta}$ due to the shift
of the injection energy by the adiabatic expansion. The other model
parameters are the same as the fiducial model. We see the spectral
hardening is prominent only for $\Delta r>2$ and there is no
hardening without adiabatic expansion, i.e. $\Delta r=1$. Because
$\gamma_{\rm inj}$ is fixed, the radio spectrum also changes
slightly with $\Delta r$. Electron injection from jet to a hotspot
AR has long been debated, the so-called injection problem of shock
models \citep[e.g.,][]{car91}.  A wide range of $\gamma_{\rm inj}$
is used by different authors, e.g. $\gamma_{\rm inj} = 10$ used by
\citet{Cro05}, $\gamma_{\rm inj} = 100$ by \citet{car91}, and
$\gamma_{\rm inj} = 1000$ by \citet{w98}.
In the context of SA,  an electron energy comparable to the proton
rest mass energy is favored because lower energy electrons can be
accelerated efficiently by whistler waves \citep{lmpf06}. A recent
reanalysis of the BBS of the hotspots in Cygnus A also suggests a
sharp spectral break at this energy \citep{s07}. We therefore choose
$\gamma_{\rm inj}=1800$ for the fiducial model. The two thin solid
lines in Figure \ref{model} show the spectra for $\gamma_{\rm inj} =
10$, and 3000.  The latter under-predicts the 90 cm radio flux by
$\sim 60\%$. Here $n_e\propto \gamma_{\rm inj}^{-\delta}$ and
$U_e/U_B\propto \gamma_{\rm inj}^{1-\delta}$.
Figure \ref{model} also shows the spectral evolution assuming a
constant injection power and constant magnetic field and electron
density in the CR. As expected, the total luminosity of the hotspot
increases with time. At $t=0.4$ Myr, the X-ray flux is dominated by
the SSC component, which may explain the observation that the X-ray
emission of high-power hotspots are dominated by SSC \citep{h04}.
The SSC component becomes more prominent if a bigger $U_e/U_B$ is
chosen for the fitting, which also results in a lower $B$, higher
$\gamma_c$, and higher energy flux from the AR.

Several models have been proposed for the X-ray spectral hardening
before. None of them appears to be applicable to WHPA. We also
consider the Klein-Nishina effects of SSC on the electron
distribution. Figure 2 shows that the volume of the emission region
needs to be at least 10 times smaller than the fiducial model to
make the SSC cooling dominant. This leads to an electron to magnetic
field energy density ratio greater than $150$ \footnote{The
synchrotron luminosity $L_{\rm syn}\propto N_0 R^3 B^2\gamma_X^2$,
where $N_0$ is the normalization of the electron distribution and
$\gamma_X$ is the Lorentz factor of electrons producing keV X-rays
through synchrotron process, i.e. $B\gamma_X^2$ is on the order of
$10^{11}$ G. To make the SSC cooling dominant, the energy density of
synchrotron photons needs to be greater than the magnetic field
energy density, i.e. $B^2 R^2 < 2 L_{\rm syn}/c$. The electron
energy density is proportional to $N_0\gamma^{1-\delta}_{\rm inj}$.
Then we have $B^2/N_0\gamma^{1-\delta}_{\rm inj}\propto B^4
R^3\gamma^2_X/L_{\rm syn}\gamma^{1-\delta}_{\rm
inj}<10^{11}(2/c)^{3/2} L_{\rm syn}^{1/2}\gamma^{\delta-1}_{\rm
inj}$ G.  One needs to increase $\gamma_{\rm inj}$, which is
constrained by radio flux densities, to bring the magnetic field and
electrons close to energy equipartition.} and significant SSC
contributions to X-ray emission. The SA itself can also produce
hardening near the high energy cutoff \citep{s84}. However, this is
significant only for $\tau_{\rm esc}\gg\tau_{\rm ac}$
($\delta\simeq0$). The corresponding electron spectral index is
$\sim1$, which is not consistent with observations.

\section{Conclusions and Discussion}

We have shown that SA reproduces the X-ray spectral hardening of the
bright core of WHPA and predicts a spectral cutoff near $1$ keV.
With few modification, the model can be applied to other {\it
Chandra} X-ray features with similar properties. We emphasize that
the shock model has difficulties in explaining these observations
since the lifetime of X-ray emitting electrons is too short for the
shock to produce the observed high flux levels. This challenges any
models with particles accelerated efficiently in very small regions
With a very large shock cross section ($>0\farcs3^2$) and high
downstream flow velocity ($>0.1c$), the shock model may fit the BBS
of WHPA by assuming a quick adiabatic expansion phase of the
downstream flow, similar to the SA model discussed above. However, a
power-law spectrum with a photon spectral index of 2.24 is expected
beyond the X-ray band except that the acceleration rate of highest
energy electrons is fine-tuned to be equal to the cooling rate of
X-ray emitting electrons. In this case the model predicted BBS will
be indistinguishable from that of the SA model. For Bohm diffusion,
the acceleration rate is equal to $eB\Delta r^2 c/2\pi$, the
emission spectrum cuts off at $\sim 70$ MeV \citep{bg70}.
Observations above the {\it Chandra} band (e.g. with {\it Suzaku})
can readily distinguish these two models.

The SA is less efficient and operates over an extended region with
its size determined by the turbulence generation and decay
processes. We predict that the observed harder X-ray emission
originates from the AR while
the radio (optical) emission comes from the CR. The energy
dissipation and the accompanying SA may still be triggered by shock
waves. In this case, the traditional study of shock dynamics can be
used to make predictions on the displacements between peaks in the
X-ray and radio (optical) images. These may explain observations of
bright knots in FR I galaxies.  The dynamics of jet flows is
sophisticated. Some of the energy dissipation regions may not be
associated with shock fronts. In this case, one may use MHD
simulations to identify regions with strong energy dissipation and
apply the SA model
to make predictions on the source morphology.
MHD simulations may also address the dynamics related to the
turbulence and the adiabatic expansion phase introduced to produce
the observed spectral hardening.

\vspace{-.4cm}\acknowledgments We thank the referee for careful
reading and very helpful comments. This work was supported in part
by the NSF of China (grants 10325313, 10733010 and 10521001), CAS
key project (KJCX2-YW-T03), the PDF of China (20070410636) and under
the auspices of the US Department of Energy by its contract
W-7405-ENG-36 to LANL.

\clearpage

\end{document}